RESEARCH

# $^{155}$Tb production by cyclotrons: what level of $^{155}$Gd enrichment allows clinical applications?

Francesca Barbaro[1,2], Luciano Canton[1], Nikolay Uzunov[3], Laura De Nardo[1,4†*] and Laura Melendez-Alafort[5†]

[1]INFN-Padua, National Institute of Nuclear Physics, Via Marzolo 8, I-35131 Padua, Italy

[2]Department of Physics, University of Pavia, Via A. Bassi 6, I-27100 Pavia, Italy

[3]INFN-Legnaro National Laboratories, National Institute of Nuclear Physics, Viale dell'Università 2, I-35020 Legnaro, Italy

[4]Department of Physics and Astronomy, University of Padua, Via Marzolo 8, I-35131 Padua, Italy

[5]Veneto Institute of Oncology IOV-IRCCS, Via Gattamelata 64, I-35138 Padua, Italy

*Correspondence: laura.denardo@unipd.it

†Co-senior authors: De Nardo and Melendez-Alafort

## Abstract

**Background**: $^{155}$Tb represents a potentially useful radionuclide for diagnostic medical applications, but its production remains a challenging problem, in spite of the fact that many production routes have been already investigated and tested. A recent experimental campaign, conducted with low-energy proton beams impinging on a $^{155}$Gd target with 91.9% enrichment, demonstrated a significant co-production of $^{156g}$Tb, a contaminant of great concern since its half-life is comparable to that of $^{155}$Tb and its high-energy $\gamma$ emissions severely impact on the dose released and on the quality of the SPECT images. In the present investigation, the isotopic purity of the enriched $^{155}$Gd target necessary to minimize the co-production of contaminant radioisotopes, in particular $^{156g}$Tb, was explored using various computational simulations.

**Results**: Starting from the recent experimental data obtained with a 91.9% $^{155}$Gd-enriched target, the co-production of other Tb radioisotopes besides $^{155}$Tb has been theoretically evaluated using the Talys code. It was found that $^{156}$Gd, with an isotopic content of 5.87%, was the principal contributor to the co-production of $^{156g}$Tb. The analysis also demonstrated that the maximum amount of $^{156}$Gd admissible for $^{155}$Tb production with a radionuclidic purity higher than 99% was 1%. A less stringent condition was obtained through computational dosimetry analysis, suggesting that a 2% content of $^{156}$Gd in the target can be tolerated to limit the dose increase to the patient below the 10% limit. Moreover, it has been demonstrated that the imaging properties of the produced $^{155}$Tb are not severely affected by this level of impurity in the target.

**Conclusions**: $^{155}$Tb can be produced with a quality suitable for medical applications using low-energy proton beams and $^{155}$Gd-enriched targets, if the $^{156}$Gd impurity content does not exceed 2%. Under these conditions, the dose increase due to the presence of contaminant radioisotopes remains below the 10% limit and good quality images, comparable to those of $^{111}$In, are guaranteed.

## Background

Terbium is one of the few rare earth radiometals that could be used in nuclear medicine for tumor diagnosis and treatment, due to the favorable physical decay properties such as half-lives, type and energy of emissions. The four Tb radionuclides with higher clinical interest are $^{152}$Tb and $^{155}$Tb, positron and gamma emitters respectively, relevant for diagnostic purposes, and $^{149}$Tb and $^{161}$Tb, $\alpha$ and $\beta^-$ emitters respectively, suitable for therapeutic applications (1). The matched pairs therapy/diagnostic radioisotopes of this element could be ideal candidates for theranostic applications, with no differences in chemical and pharmacokinetics behaviours (2). Indeed, significant differences could be observed when using pairs of diverse elements for diagnostics and therapy, because it is well known that each metal ion has specific chemical demands arising from its fundamental characteristics, giving rise to different coordination and geometry numbers (3), which influence the pharmacological properties of the labelled molecules.

Positron emission tomography (PET) imaging is the nuclear imaging technique with higher sensitivity and resolution. Although, single photon emission computed tomography (SPECT) is still the predominant technology worldwide, due to the easy availability of $\gamma$-emitter radionuclides and the low cost of SPECT camera compared to PET scanner (4). Therefore, the production of new gamma-emitting radionuclides that could be detected with the large number of available SPECT scanners is encouraged. Among Tb isotopes, $^{155}$Tb is the only suitable for SPECT imaging, thanks to its two $\gamma$ emissions at 87 keV (32%) and 105 keV (25%). Preclinical studies in nude mice bearing tumor xenograft demonstrated that $^{155}$Tb-labelled biomolecules were able to visualize tumor sites using a small-animal SPECT/CT scanner even 7 days after radiocomplexes administration (5). $^{155}$Tb has been produced by different methods although most of them generated a large number of isotopes. First investigations on the production of $^{155}$Tb via proton-induced reactions on Gd targets date back to 1989 with the work by Dmitriev *et al.* (6). In this case, no cross section measurements were performed, but rather irradiation experiments of $^{nat}$Gd thick targets for yields and activities, with incident proton energies between 11 and 22 MeV. Using $^{nat}$Gd targets, cross-section measurements were performed much later by Vermeulen *et al.* (7). Besides the detailed experimental work that included the measurement of cross sections for the formation of a variety of terbium radionuclides, theoretical calculations were also performed for all contributing reactions. Since natural Gd ($^{nat}$Gd) has seven isotopes it is not suited as target material for radioisotope production with high radionucludic purity (RNP). Therefore, comparisons were made with accurate theoretical simulations in order to identify the most promising reactions that could be effectively employed in studies with enriched targets (7). More recently Formento-Cavaier *et al.* (8) extended the measurements with $^{nat}$Gd targets up to 70 MeV. Experimental investigations with enriched targets of $^{155}$Gd and $^{156}$Gd have been undertaken by Favaretto *et al.* (9), with the result that $^{156}$Gd(p,2n)$^{155}$Tb provides high production yields, but implies the use of higher-energy cyclotrons and a significant contamination by the co-production of $^{156}$Tb. The $^{155}$Gd(p,n)$^{155}$Tb reaction, on the other hand, can be performed with medical cyclotrons, with incident protons up to around 18 or 20 MeV, and has the potential to provide significant yields with high purity. In a subsequent publication Dellepiane *et al.* (10) used the same enriched gadolinium oxide ($^{155}$Gd 91.9% enrichment or $^{156}$Gd 93.3% enrichment) as target materials, measuring the $^{155}$Tb production cross sections as well as a variety of contaminants produced for such highly enriched targets. In the most favorable case of the $^{155}$Gd target, using an input energy of 10.5 MeV, contamination from $^{156}$Tb was still significant, leading to a maximum RNP not greater than 93% after a decay time of about 96 h. Thus, it was suggested, as possible solution, to purify the final product through an off-line mass separation technique at the expense of a lower production yield, due the low efficiency of the current mass separation approaches. The production of $^{155}$Tb and other Tb-isotopes ($^{149}$Tb, $^{152}$Tb and $^{161}$Tb) has been explored also at CERN-ISOLDE using spallation of high-energy p-beams on Tantalum targets, followed by ionization and mass separation (1,11). The drawback stands in the limited quantity of $^{155}$Tb that can be produced with this method, due to the observed low efficiency of the accumulation procedure (12,13). It has been suggested also to irradiate a $^{159}$Tb target by intermediate energy (60, 70 MeV) p beams (14,15). This would open the possibility to have a $^{155}$Dy/$^{155}$Tb generator system, similarly to the renowned $^{99}$Mo/$^{99m}$Tc one. However, the double separation chemistry among lanthanides represents a crucial step that still needs to be solved and the possible co-production of very long-lived terbium contaminants could imply a too low specific activity.

In this work Talys calculations of $^{155}$Tb cross section have been benchmarked with data measured by Dellepiane *et al.* (10) with an enriched $^{155}$Gd target. The contribution of each isotopic component has been disentangled to expose the effects deriving from the impurities of the target. Once the modeling has been tested on the specific isotopic $^{155}$Gd abundance of the enriched target used by Dellepiane *et al.*, the level of enrichment of the $^{155}$Gd target necessary to produce $^{155}$Tb with high purity has been investigated. However, the assessment of RNP is not enough to evaluate the suitability of the produced $^{155}$Tb for clinical purposes.

In fact, different produced contaminants can have a different impact on the dose delivered to the patient and on the quality of the SPECT images. Therefore, biodistribution data of the DOTA-folate conjugate $^{161}$Tb-cm09 in IGROV-1 tumour-bearing mice (11) has been used to assess the dose increase (DI) to the patient following $^{155}$Tb-cm09 injection, due to the presence of contaminants in the $^{155}$Tb produced supposing different enrichment of the $^{155}$Gd target. Moreover, the effect of the contaminants with high-energy $\gamma$-emissions on the SPECT images quality was evaluated.

## Methods

### Cross sections and thick target irradiation

The study of the nuclear reaction routes implies the adoption of different models to consider both the compound nucleus formation/decay and pre-equilibrium dynamics. To this purpose the Talys code (16) has been used, specifically version 1.95 which includes the geometry dependent hybrid (GDH) model (17,18). To describe the nuclear reaction mechanisms this Talys version provides 5 pre-equilibrium (PE) and 6 level-density (LD) models, for a total of 30 possible combinations of models. The description of the PE processes is mainly based on the exciton model. Transition matrices are the basic building blocks of the exciton formalism and they are described 1) analytically; 2) numerically; or 3) derived from the imaginary component of the optical potential. A fully quantum-mechanical approach, based upon the Feshbach-Kerman-Koonin theory and alternative to the exciton model, is included and labelled as 4). The already mentioned GDH model, which considers nuclear surface effects in the exciton model, is denoted as option 5). Another important ingredient for the compound nucleus formation is the nuclear LD, which affect the cross section through the Hauser-Feshbach formalism. Talys considers three phenomenological LD models, namely 1) the Fermi gas with constant temperature; 2) the back-shifted Fermi gas; 3) the generalized superfluid model; and three microscopic models, 4) the Goriely's tabulated Hartree-Fock densities; 5) the Hilaire's tabulation based upon the combinatorial model, and 6) the Hartree-Fock-Bogoliubov temperature-dependent formalism. The variability of these models can be described statistically introducing an interquartile band, measuring the model dispersion between the lower Q1 and the upper Q3 quartile, as has been discussed in a recent paper (19). In addition, the modeling has been tested on the specific isotopic abundances of the enriched $^{155}$Gd target used by Dellepiane *et al*.

The modeled cross sections have been employed to evaluate the radionuclides produced under specific irradiation conditions, including the level of enrichment of the target. The computational approach discussed by Canton *et al.* (20) has been used for the evaluation of rates, activities, yields, and purities.

The rate *R* of production of a radionuclide from a beam colliding on a thick target can be derived from the expression

$$R = \frac{I_0}{z_{proj}|e|} \frac{N_a}{A} \int_{E_{out}}^{E_{in}} \sigma(E) \left( \frac{dE}{\rho_t dx} \right)^{-1} dE \quad (1)$$

where $I_0$ is the beam current, $z_{proj}$ is 1 for a proton beam, $e$ the electron charge, $N_a$ the Avogadro number, $A$ the atomic mass of the target element, $E_{in}$ and $E_{out}$ the energy of the beam hitting the target and the one leaving the target after traveling through its thickness, respectively. The production cross section for the nuclide is $\sigma(E)$, while the target density $\rho_t$ and the stopping power of the projectile in the target dE/dx, described by the Bethe-Bloch formula (21).

Once the rates given in eq. (1) for all the Tb radionuclides of interest are determined, it is possible to calculate, by means of standard decay treatment of the isotopes via the Bateman equations, the time evolution of the number of produced radionuclides, and from there the evolution of the activities during and after an irradiation. The time dependence of the number of Tb radionuclides produced and their corresponding activities are used to evaluate the isotopic and radionuclidic purities for the production of $^{155}$Tb. The main decay data of interest for the present analysis are reported in Table 1.

**Table 1.** Main decay data of $^{xxx}$Tb radionuclides.

| Radionuclide | Decay mode (%) | $T_{1/2}$ | Emitted energy (MeV/nt) | | |
|---|---|---|---|---|---|
| | | | Electron | Photon | Total |
| $^{154g}$Tb | EC, $\beta^+$ (100%) | 21.5 h | 0.0681 | 2.2831 | 2.3512 |
| $^{154m1}$Tb | EC (78.20%), IT (21.80%), $\beta^-$ (< 0.10%) | 9.4 h | - | - | - |
| $^{154m2}$Tb | EC (98.20%), IT (1.80%) | 22.7 h | - | - | - |
| $^{155}$Tb | EC (100%) | 5.32 d | 0.0434 | 0.1777 | 0.2211 |
| $^{156}$Tb | EC (100%) | 5.35 d | 0.0835 | 1.9371 | 2.0206 |
| $^{156m1}$Tb | IT (100%) | 24.4 h | 0.0171 | 0.0370 | 0.0540 |
| $^{156m2}$Tb | IT (100%) | 5.3 h | 0.0874 | 0.0048 | 0.0922 |

Decay mode, half-life, emitted energy per nuclear transformation (nt) in the form of electron, photon, and the total one of $^{xxx}$Tb radionuclides were obtained by the ICRP 107 publication (22) and NUDAT3 database (23) (emitted energy data not available for $^{154m1}$Tb and $^{154m2}$Tb).

## Assessment of organ absorbed doses and effective dose due to $^{xxx}$Tb-cm09 injection

Dosimetric assessment were carried out using biodistribution data of the DOTA-folate conjugate $^{161}$Tb-cm09 in IGROV-1 tumour-bearing mice. The cm09 is composed by a targeting vector (which selectively binds to the folate receptor expressed on a variety of tumour types) conjugated to small-molecular-weight albumin (which improves the blood circulation time and tissue distribution profile of folate conjugates) and to the DOTA chelating agent. Dosimetric evaluation has been performed supposing the cm09 labelled with $^{155}$Tb and also with other Tb-radioisotopes expected to be produced by proton irradiation of $^{155}$Gd-enriched targets.

Biodistribution data of $^{161}$Tb-cm09 acquired in a time window of 7 days post injection in IGROV-1 tumour-bearing female nude mice (11) were used to estimate the absorbed doses to humans due to different Tb-radioisotopes through the relative mass scaling method, which takes into account the differences in human and animal organ masses compared to the total body masses (24). The activity concentrations in the different animal source organs (blood, lung, spleen, kidneys, stomach, intestines, liver, salivary glands, muscle and bone), reported as per cent of injected activity per gram of tissue ($[\%IA/g]_A$), were scaled from mice to adult male humans to obtain the decay-corrected per cent of injected activity for each human source organ ($[\%IA/organ]_H$) through the following formula:

$$\left(\frac{\%IA}{organ}\right)_H = \left(\frac{\%IA}{g}\right)_A \cdot \frac{OW_H}{TBW_H} \cdot TBW_A \qquad (2)$$

with $OW_H$ the weight of human organ, $TBW_A$ and $TBW_H$ the total body weight for animal and human, respectively. $OW_H$ and $TBW_H$ values were obtained from the adult male phantom implemented in the Organ Level Internal Dose Assessment (OLINDA) software code (25). Biodistribution data were then plotted as a function of post injection (p.i.) time and fitted by a tri-exponential equation, representing the phase of accumulation and the possibility of both a fast and a slow elimination of the radiopharmaceutical, with CoKiMo software (26). At last, the number of disintegrations per unit of administered activity in the source organs was obtained by integration of organ activity curves for each $^{xxx}$Tb-radioisotope, considering its physical half-life. These data were then used as input values in an adult male phantom to perform dosimetric calculations with the OLINDA software code version 2.2.3 as reported before (27), obtaining the organ absorbed doses per unit of administered activity. For each $^{xxx}$Tb radioisotope, the specific effective dose, $^{xxxTb}ED$ corresponds to the sum of the product of the organ equivalent dose per unit of administered activity, $^{xxxTb}D_{org}$, and the respective tissue-weighting factor, $w_{org}$, recommended by ICRP 103 (28),

$$^{xxxTb}ED = \sum_{org} {^{xxxTb}D_{org}} \cdot w_{org}$$

(3)

The irradiation simulations with different target enrichment leads to a distinct distribution of $^{xxx}$Tb activities, described by the fraction of total activity

$$f_{^{xxx}Tb}(t) = \frac{A_{^{xxx}Tb}(t)}{\sum_{yyy} A_{^{yyy}Tb}(t)} \tag{4}$$

The total effective dose ($ED_{tot}$) of Tb-cm09 were calculated at different times after the end of bombardment (EoB):

$$ED_{tot}(t)=\sum_{xxx} f_{^{xxx}Tb}(t)\,ED^{^{xxx}Tb} \quad (5)$$

Finally, the evaluation of the DI generated by the co-produced Tb-impurities is determined according to the following equation

$$DI(t)=\frac{ED_{tot}(t)}{ED^{^{155}Tb}}$$

(6)

and represents the ratio between the total effective dose and the effective dose due to an ideal injection of pure $^{155}$Tb compound. The doses discussed above refer to the total doses of an injection delivered at time *t* post-production.

**Assessment of the imaging properties of $^{155}$Tb in the presence of other contaminant isotopes**

The main problem in the quality of SPECT images when $^{155}$Tb contains γ-emitting contaminants is the noise produced by the Compton scattering of high energy gamma rays. Even using energy windows to acquire only the gamma rays of interest, higher energy γ-rays can be Compton scattered and registered by the acquisition system. The effect of such signals results in blurry and bad contrast images. Therefore, a study of the noise contribution of high-energy gamma-rays emitted by contaminant terbium isotopes has been carried out using a preclinical PET/SPECT/CT system (VECTor 5, Milabs) as a model of the acquisition system.

To simulate the γ-ray spectrum of each Tb radionuclide in VECTor 5, a homemade menu-driven spreadsheet software Visual Gamma was used. It makes use of specific initial system parameters such as efficiency and peak shape and simulates the spectrum of the radionuclides, using as input data the energy and the intensity of the γ-ray, taken from the NUDAT3 database (23). Parameters such as γ-ray efficiency, the peak width and the shape of the Compton area as a function of the γ-rays energies have been determined using standard calibration point-like sources ($^{57}Co$, $^{22}Na$, $^{60}Co$ and $^{137}Cs$) in air. The process of assessment of the image quality, obtained using a particular γ-ray peak, was based upon two main factors: the intensity of the γ-ray emission and the presence of Compton scattered γ-rays in the energy window of the peak selected for imaging. An energy window of 10% interval around the peak barycenter has been chosen. To evaluate the influence of Compton scattered γ-rays on the final quality of the image, the ratio $\delta = N_{extr}/N_{intr}$ was used, where $N_{extr}$ is the number of all higher-energy γ-rays that underwent Compton scattering and fell into the selected energy window, $N_{intr}$ is the number of γ-rays from the selected photo-peak that fell into the energy window, i.e. not including Compton scattering contribution. It is convenient to use such Compton-to-peak ratio since in the ideal case (lack of other noises) its value directly corresponds to the amount of the noises in the image.

Imaging qualities were evaluated for $^{155}Tb$ produced from the proton irradiation of $^{155}Gd$ targets with enrichment of 100, 99 and 98%, at the EoB and 96 h later.

## Results

### Cross sections

The same composition of the enriched $^{155}Gd$ targets employed by Dellepiane (10) for cross section measurements was used in the irradiation simulations and specifically, $^{154}Gd$ 0.5%, $^{155}Gd$ 91.90%, $^{156}Gd$ 5.87%, $^{157}Gd$ 0.81%, $^{158}Gd$ 0.65%, $^{160}Gd$ 0.27%. In Fig. 1 the experimental cross sections for $^{155}Tb$ production are compared with theoretical calculations. The gray band represents the models variability between Q1 and Q3, the dashed lines are the minimum/maximum values of the cross sections obtained with all models, and the blue solid line is the Talys default option (PE2-LD1), commonly referred to as the standard simulation in the Literature. For the $^{155}Tb$ case all models reproduce equally the cross section up to 10 MeV, and for higher energy the band is relatively thin and this corresponds to a limited model variability, as expected from a typical (p,n) reaction. On account of this, and because the Talys default reproduces the very recent data measured by Dellepiane (10) quite satisfactorily, the analysis will be presented considering the Talys default as the benchmark calculation.

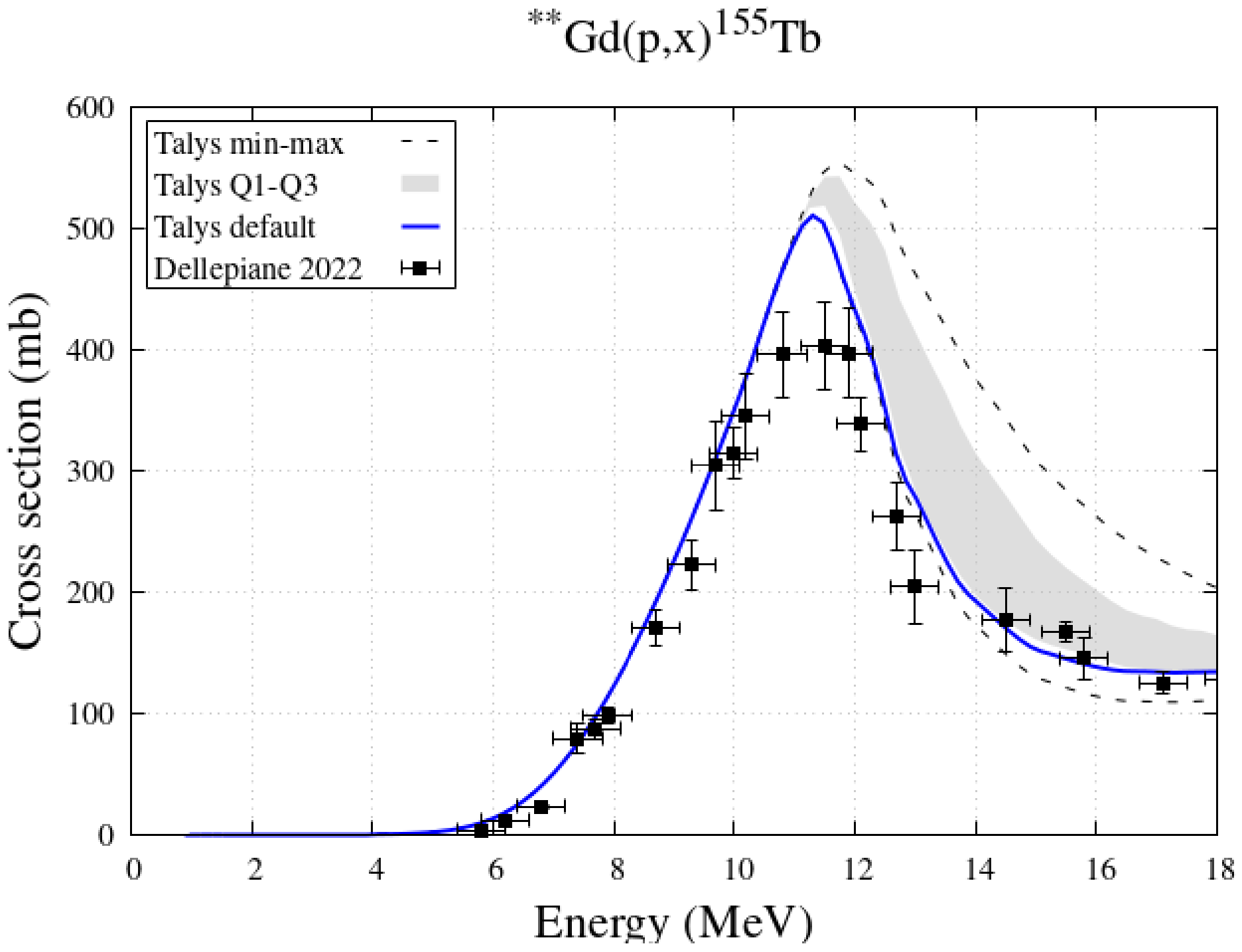

**Fig. 1** Experimental $^{155}Tb$ cross sections from enriched $^{155}Gd$ target and theoretical curves expressing the variability of nuclear reaction models.

Fig. 2 shows the contribution of each isotopic component of the target to the $^{155}Tb$ production. The main contribution to the $^{155}Tb$ cross-section comes, as should be expected, from the (largest) $^{155}Gd$ component of the target. For energies higher than 10 MeV, also the contribution from the $^{156}Gd$ component becomes significant, and this explains the increase of the experimental data with respect to an ideal target with 100% $^{155}Gd$ enrichment. The contribution from $^{157}Gd$ is quite small and that from other Gadolinium components of the targets, such as $^{154}Gd$ is negligible.

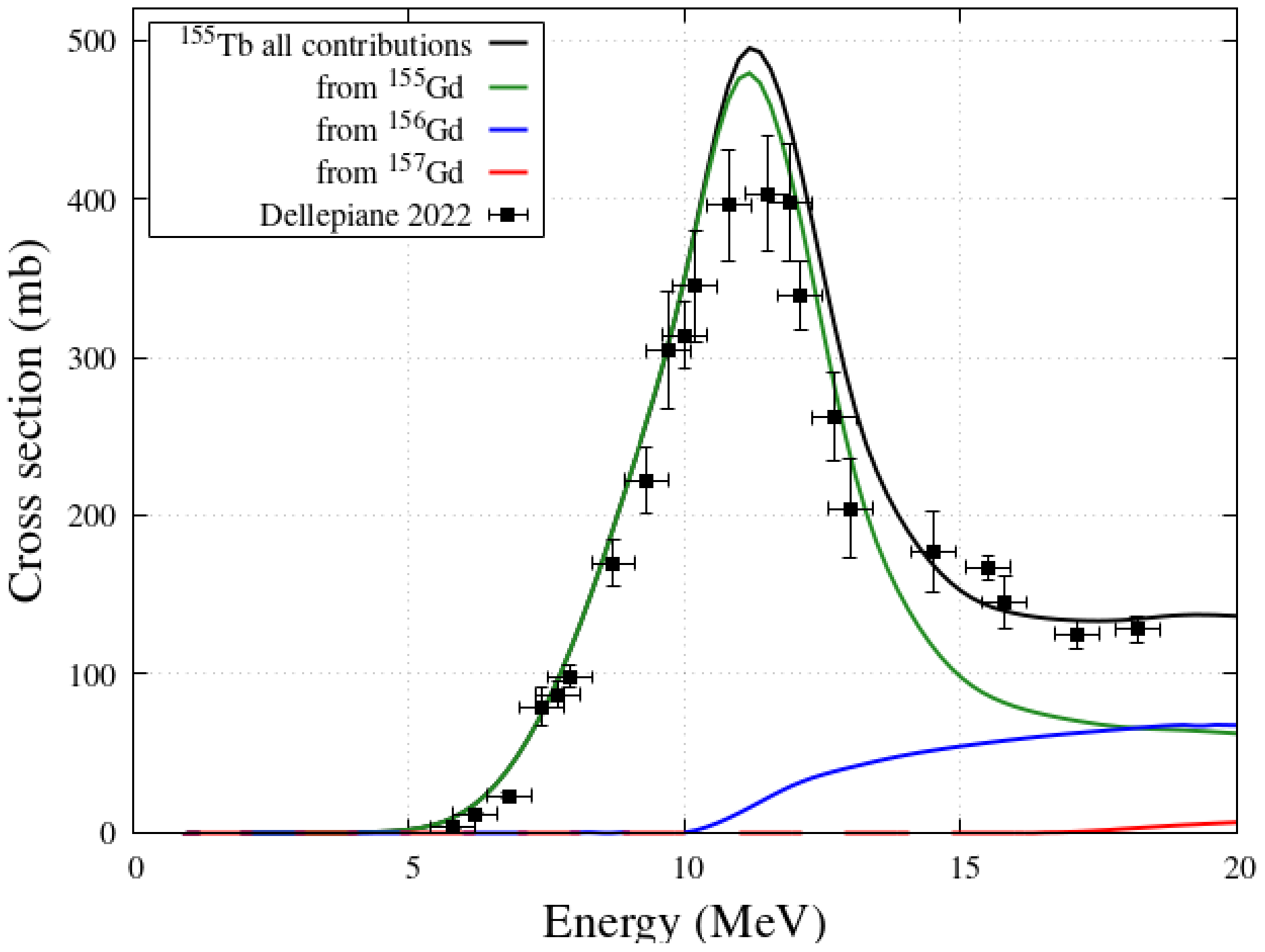

**Fig. 2** $^{155}$Tb cross sections from enriched $^{155}$Gd target: Talys calculations and contributions from different Gd isotopes of the target

.Next, the $^{156t}$Tb total cross-section, which refers to the cumulative cross sections of $^{156g,\ 156m1,\ 156m2}$Tb (ground and the first two metastable states), is presented in Fig. 3 A. The curve including all contributions is in agreement with the data measured by Dellepiane *et al*. It is evident that the main contribution belongs to the $^{156}$Gd component of the target, followed by the $^{157}$Gd component at slightly higher energies. At even higher energies, about 19 MeV, the contribution from $^{158}$Gd also becomes significant. Instead, the contribution to the $^{156}$Tb cross section from the main $^{155}$Gd component of the target remains quite small in the entire range of considered energies.

Fig. 3 B, C and D describe the $^{156}$Tb production cross sections for ground and the two metastable states, respectively. For the ground state, the model calculations are in fair agreement with the measurements, although at higher energies there is a tendency to underestimate the data. For the first metastable state there are no measured production data, while for the second one the curves underestimate the measurements. In all cases, by separating in the Gd target the individual isotopic components, it is evident that the main contributions to the production of the contaminants derive from the Gd isotopes heavier than 155.

The remaining contaminants that may have an impact on the quality of the produced $^{155}$Tb are $^{153}$Tb and $^{154}$Tb (separated in ground and the first two metastable states) and their cross sections are given in Fig. 4.

The theoretical curves are in clear agreement with the measurements, with the exception of $^{154m2}$Tb where slight discrepancies can be observed. In all figures it is evident that the contamination at energies lower that 10 MeV can be entirely ascribed to the small presence of $^{154}$Gd in the target, responsible of a characteristic bump seen in the lower energy data. At higher energies, $^{155}$Gd becomes the principal contributor to the production of these contaminants.

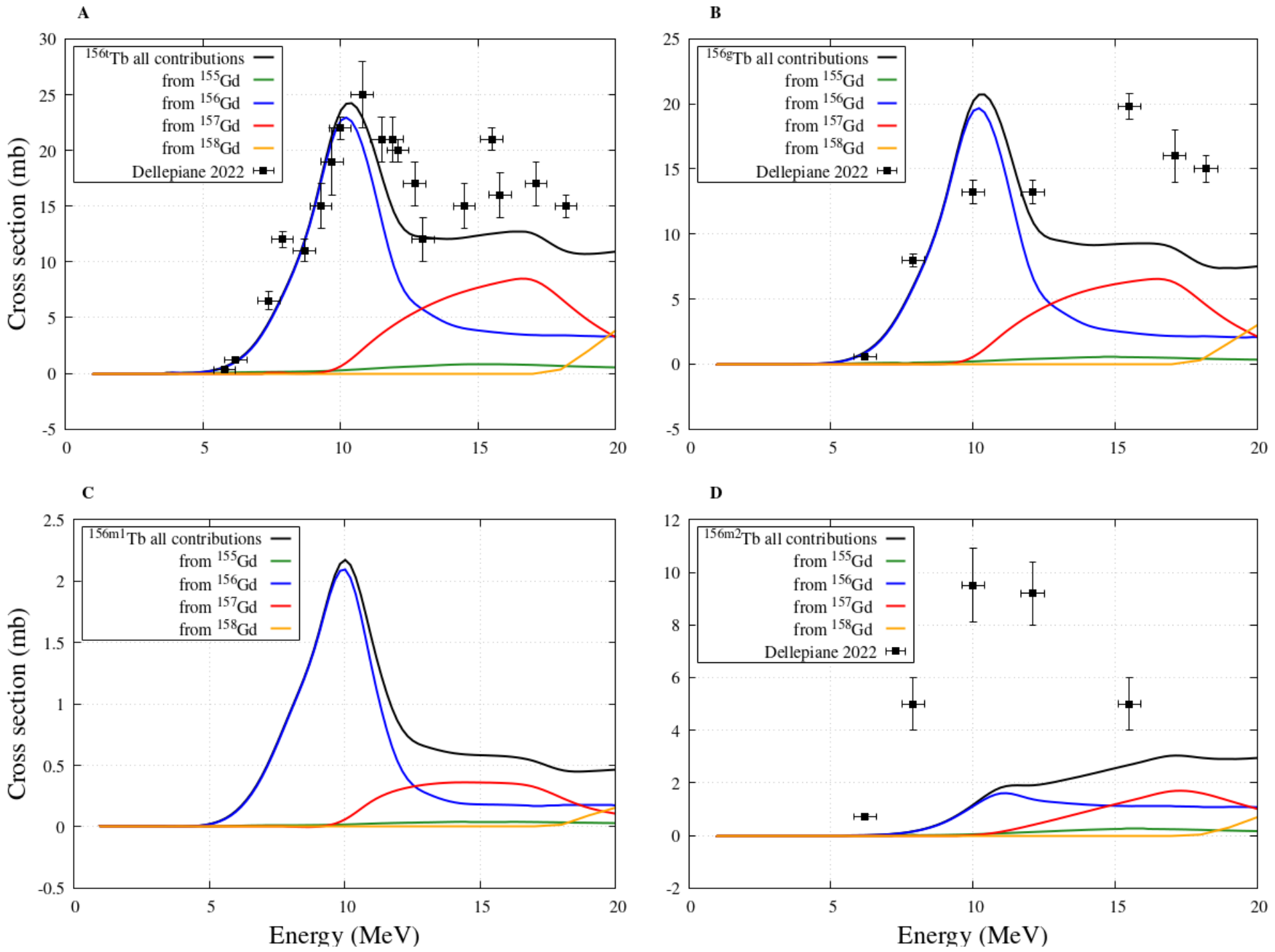

**Fig. 3** $^{156g,\ 156m1,\ 156m2,\ 156t}$Tb (where t in panel A is the cumulative cross section of $^{156g,\ 156m1,\ 156m2}$Tb) cross sections from enriched $^{155}$Gd target obtained from Talys calculations: contributions from different Gd isotopes of the target.

| Target enrichment | $^{155}$Tb | $^{156g}$Tb | $^{156m1}$Tb | $^{156m2}$Tb | $^{154g}$Tb | $^{154m1}Tb$ | $^{154m2}Tb$ |
|---|---|---|---|---|---|---|---|
| $^{155}$Gd-100% [EoB] | 4.383 | 0.0027 | 0.0013 | 0.0121 | 0.0293 | 0.121 | 0.0 |
| $^{155}$Gd-99% [EoB] | 4.3424 | 0.03743 | 0.0217 | 0.05467 | 0.0290 | 0.120 | 0.0 |
| $^{155}$Gd-98% [EoB] | 4.3021 | 0.07209 | 0.04214 | 0.0972 | 0.0287 | 0.119 | 0.0 |
| $^{155}$Gd-91.90% [EoB] | 4.04829 | 0.2123 | 0.12422 | 0.2682 | 0.07016 | 0.274 | 0.008 |
| $^{155}$Gd-100% [72 h] | 2.9648 | 0.00239 | 0.00017 | 9.85E-07 | 0.00479 | 0.0006 | 0.0 |
| $^{155}$Gd-99% [72 h] | 2.9375 | 0.02977 | 0.00281 | 4.4496E-06 | 0.00475 | 0.0006 | 0.0 |
| $^{155}$Gd-98% [72 h] | 2.910 | 0.0571 | 0.0054 | 7.914E-06 | 0.004698 | 0.0006 | 0.0 |
| $^{155}$Gd-91.90% [72 h] | 2.73853 | 0.16776 | 0.01144 | 2.183E-05 | 0.01125 | 0.0013 | 0.0009 |
| $^{155}$Gd-100% [96 h] | 2.6026 | 0.002118 | 8.63E-05 | 4.26905E-08 | 0.00224 | 0.0001 | 0.0 |
| $^{155}$Gd-99% [96 h] | 2.57866 | 0.02639 | 0.0014 | 1.928E-07 | 0.00222 | 0.0001 | 0.0 |
| $^{155}$Gd-98% [96 h] | 2.5547 | 0.05067 | 0.00276 | 3.4296E-07 | 0.002196 | 0.0001 | 0.0 |
| $^{155}$Gd-91.90% [96 h] | 2.40399 | 0.14878 | 0.00812 | 9.4601E-07 | 0.00526 | 0.0002 | 0.0004 |

**Table 2.** $^{xxx}$Tb radioisotopes yields (MBq/μA·h) for different $^{155}$Gd-enriched targets at the EoB, 72, and 96 h after.

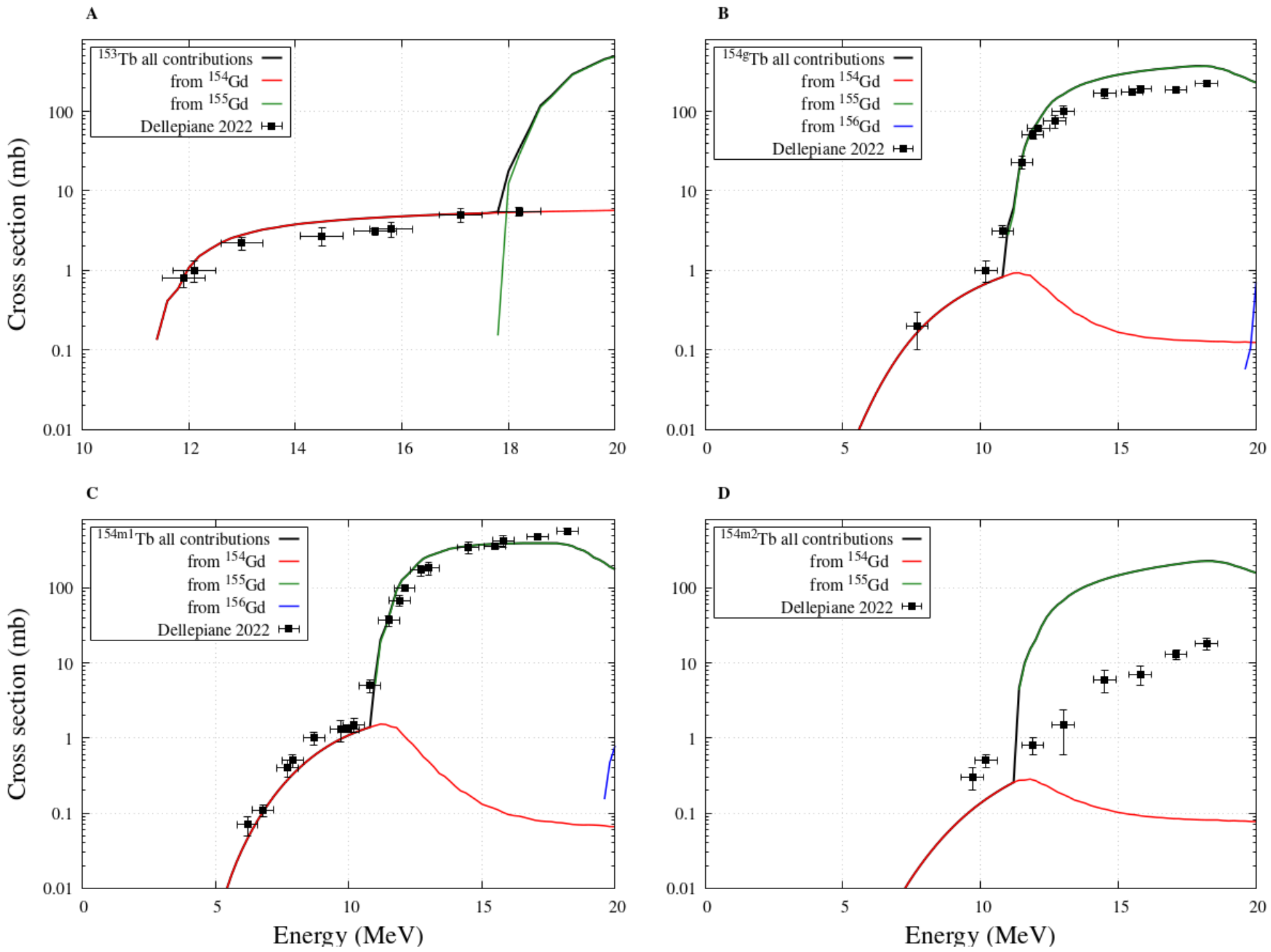

**Fig. 4** $^{153,\ 154g,\ 154m1,154m2}$Tb cross sections from enriched $^{155}$Gd target obtained from Talys calculations: contributions from different Gd isotopes of the target.

## Yields, isotopic, and radionuclidic purity

Starting from the cross sections analysis of the effects of the isotopic components of the target used for $^{155}$Tb production, yields, isotopic and RNP have been assessed for different target compositions. Four different targets were considered in the simulations to evaluate the minimum enrichment needed for a $^{155}$Tb production with the purity needed for medical applications: the first target with the exact isotopic composition considered in measurements by Favaretto *et al*. (9) and Dellepiane *et al*. (10) (in particular with 91.90% of $^{155}$Gd and 5.87% of $^{156}$Gd), two enriched targets (99% and 98% of $^{155}$Gd), considering $^{156}$Gd as the only contaminant, and the ideal target with 100% of $^{155}$Gd. The irradiation conditions, for all cases, are set to 1 $\mu$A current, 1 h irradiation time and 10.5-8 MeV energy interval, corresponding to the optimal energy selection defined by Dellepiane.

Table 2 reports the yields of the main Tb radionuclides involved in the production. The $^{156}$Tb contamination grows proportionally with the fraction of $^{156}$Gd in the target. At these energies, production of $^{153}$Tb is negligible in all cases. The $^{154g,\ 154m1,\ 154m2}$Tb contamination remains small and stable at varying the $^{156}$Gd component in the target. It increases only if the target contains a fraction of $^{154}$Gd, as is the case for the target employed by Dellepiane, with a 0.5% contribution. Table 2 also exhibits the yields 72 h and 96 h after EoB, when the activities of $^{154g}$Tb and all metastable states are considerably reduced, since their half-lives are about 1 d or less. Figure 5 shows the time evolution of the $^{155}$Tb RNP considering the four different target enrichment. The solid green line, with the larger $^{156}$Gd contamination (5.87%), settles around 93.5% in agreement with the level of RNP measured after 96 h from EoB (10). Significantly higher values are reached in the other three cases. In particular, 97.8, 98.8, and 99.8% RNP is obtained, after 96 h, with a target enrichment of 98, 99, and 100%, respectively. It is evident that the contamination of the target with $^{156}$Gd directly affects the $^{155}$Tb RNP, so it is crucial to limit it as much as possible.

Figure 6 shows the fraction of total activity of $^{154g}$Tb, $^{156g}$Tb, $^{156m1}$Tb, and $^{156m2}$Tb. Clearly, the amount of $^{156}$Gd in the target proportionally influences the activity of all three states of $^{156}$Tb. However, the main problem is represented by $^{156g}$Tb because its long half-life is comparable to the one of $^{155}$Tb. Conversely, the production of $^{154g}$Tb represents a minor issue, because of its shorter half-life. Moreover, in the selected energy region, $^{154g}$Tb is not produced by $^{156}$Gd, while it is produced by $^{154}$Gd which appears, with a residue of 0.5%, only in the less enriched target.

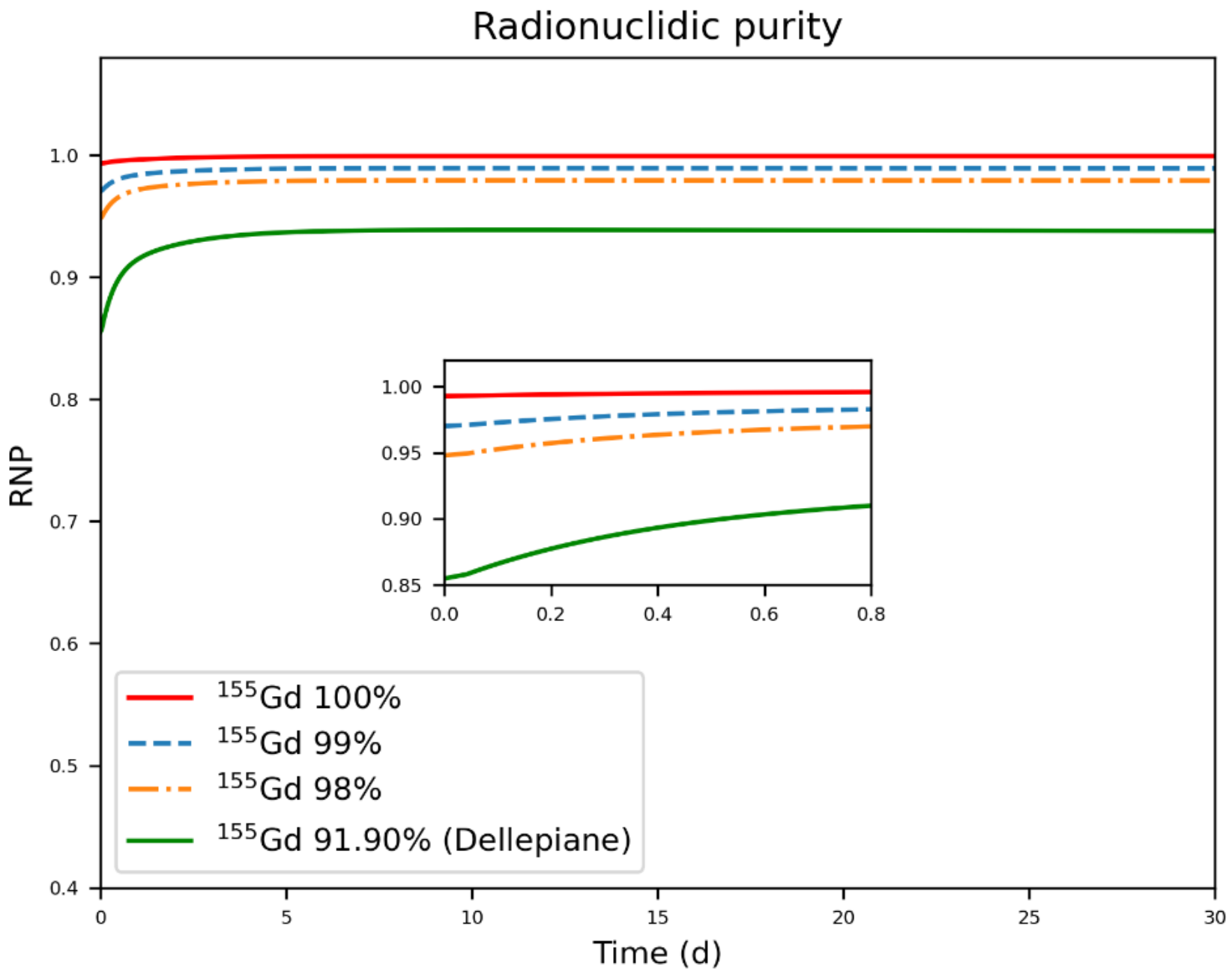

**Fig.5** $^{155}Tb$ radionuclidic purity for the different target enrichments.We have evaluated in addition the isotopic purities and, after 96 h from EoB, they correspond to 93.9, 97.9, 98.9, and 99.9% for the target enrichment of 91.90, 98, 99, and 100% respectively. This implies that the production route is essentially carrier-free, without the presence of contaminants, including those stable or with long half-lives.

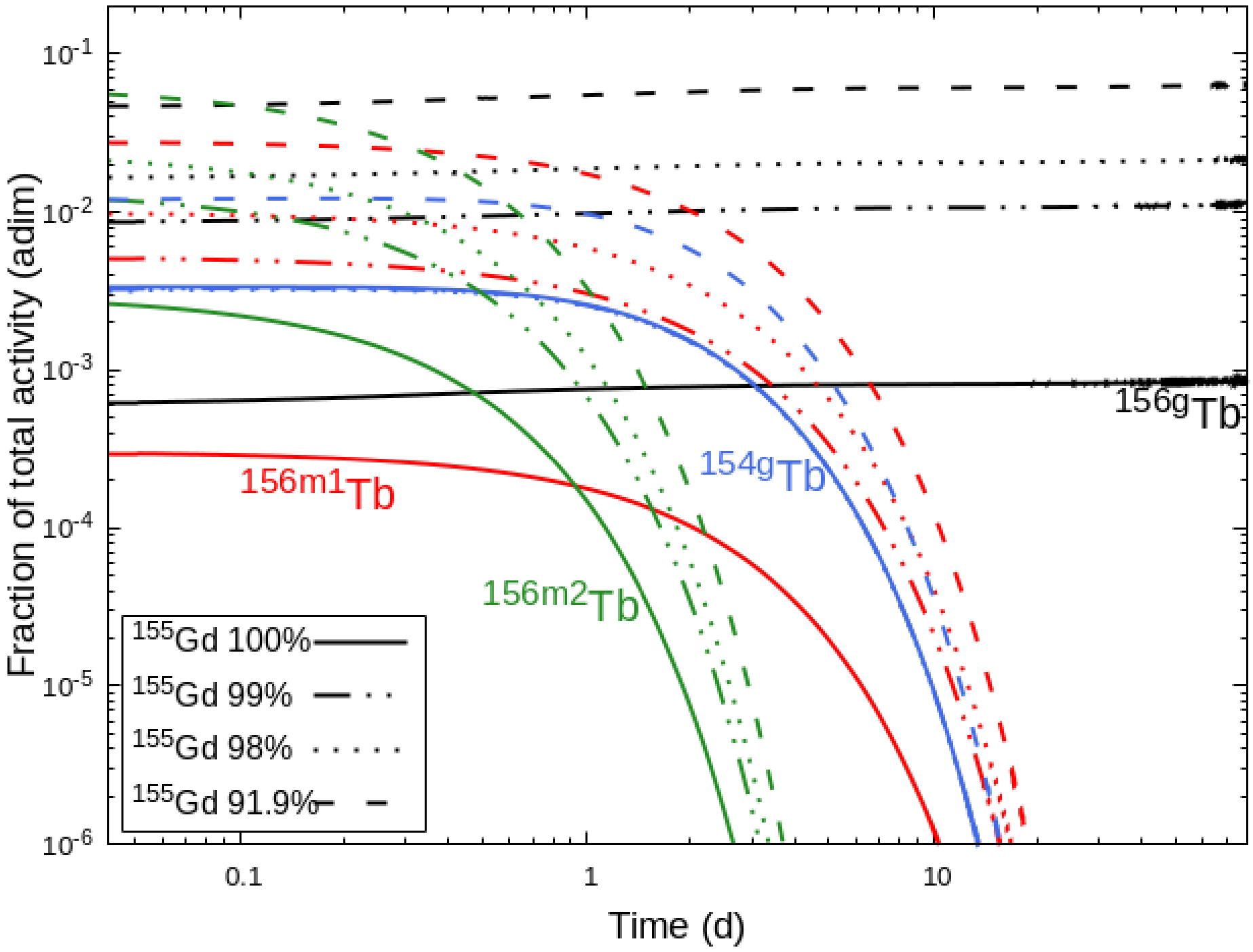

**Fig. 6** Fraction of total activity for the main contaminants $^{154g}$Tb (blue curves), $^{156g}$Tb (black curves), $^{156m1}$Tb (red curves), $^{156m2}$Tb (green curves), for different enrichment of the $^{155}$Gd targets.

## Organ absorbed doses and effective dose due to $^{XXX}$Tb-cm09 injection

Biokinetics curves were obtained plotting the radiopharmaceutical concentration corrected by the radioactive decay vs. time for each source organ of the ICRP 89 male phantom of 73 kg (29). The total volume of the blood (5110 ml) was obtained using the specific volume value of 70 ml/kg. Figure 7 shows a fast blood clearance followed by a quick radiopharmaceutical uptake by the main organs, with a slow wash-out. Liver and kidneys were the organs with slower clearance.

The number of disintegrations in the source organs, calculated by assuming that the injected radiopharmaceutical was labelled with only one of the radioisotopes $^{154g}$Tb, $^{155}$Tb, $^{156}$Tb, $^{156m1}$Tb and $^{156m2}$Tb, are reported in Table 3. These are the main radionuclides expected to be produced through the proton irradiation of the enriched $^{155}$Gd targets. The dosimetric properties of $^{154m1}$Tb and $^{154m1}$Tb were not assessed because these two metastable states are not included in the OLINDA software. However, at irradiation energy of 10.5 MeV, $^{154m1}$Tb and $^{154m2}$Tb production is essentially due to the presence of $^{154}$Gd in the target, not considered in the case of $^{155}$Gd target enrichment ≥98%. Besides, since the energy of these two metastable states is very close to that of the ground state, their contribution to the absorbed dose occurs mainly through the decay of the ground state, properly taken into account through application of the Bateman equations.

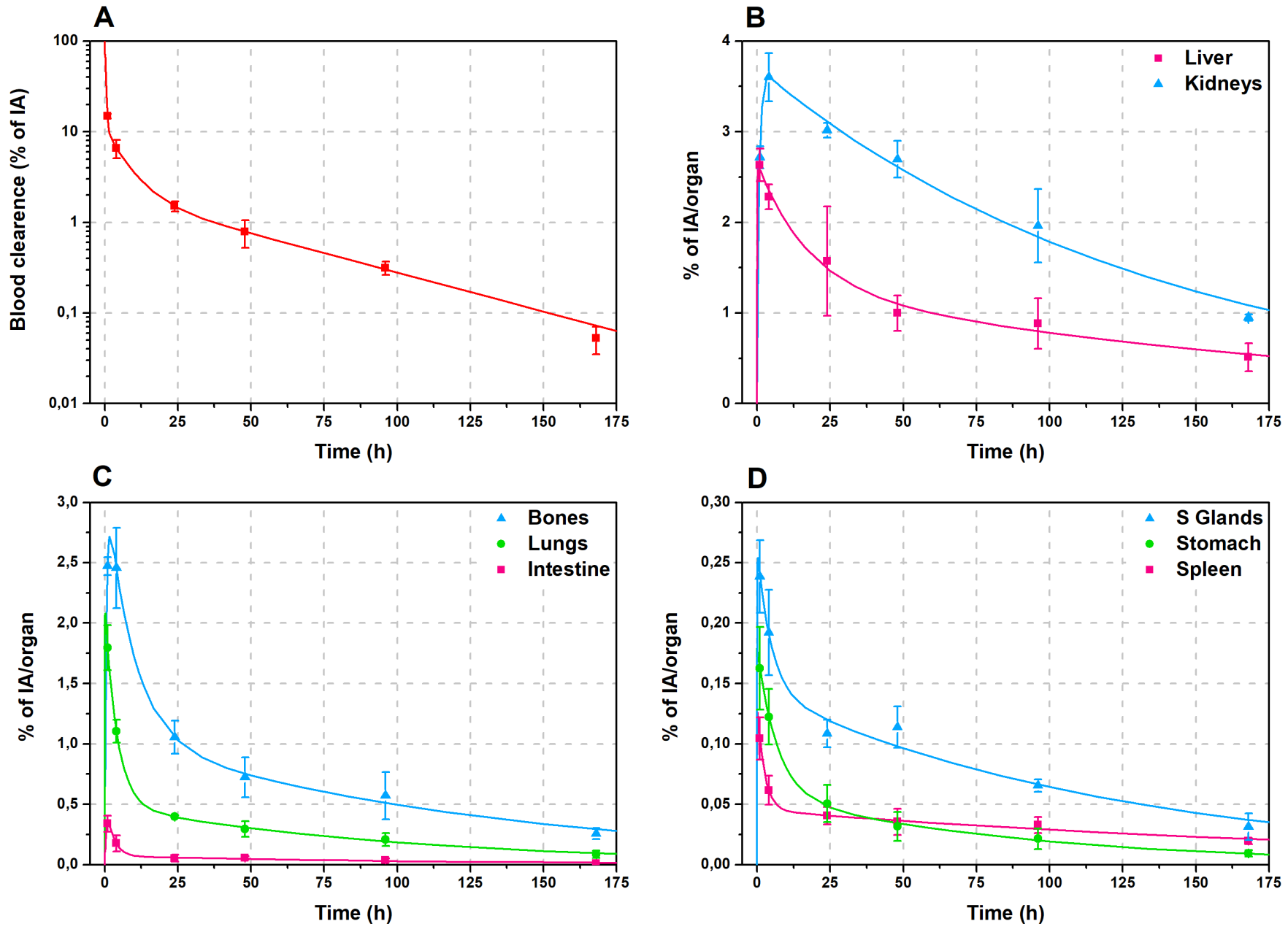

**Fig. 7.** Time-activity curves of Tb-cm09 in the main source organs: symbols show experimental data obtained from biodistribution studies; the lines depict fitted time-activity curves.

For the dosimetric assessment, the total activity in the intestines was distributed in left colon, small intestine, right colon and rectum, according to their mass ratio with respect to the total intestine mass. The mean maximum volume of blood that could be contained in the four chambers of the heart, two atria and two ventricles, of an adult man is about 505 ml (30), therefore, just the 10% of the blood activity was assigned to the "heart contents" and the rest to "Remaining". The activity in the mice muscle was extrapolated to human considering that the human muscle is the 40% of the total body weight. Muscle activity was also assigned to the 'Remaining' organs when performing calculations with OLINDA 2.2.3, because this tissue is not included in the source organs of the ICRP 89 phantom model.

For each $^{xxx}$Tb-radiopharmaceutical, the organs with the highest number of disintegrations were the kidneys, followed by the liver and the cortical bone. Comparing the different radioisotopes, it was found that the radionuclides with the highest number of disintegrations were $^{155}$Tb and $^{156g}$Tb, due to their long half-life. $^{156m2}$Tb showed the lowest number of disintegrations because of its relatively short half-life.

The absorbed doses per unit of administered activity in the main human male organs calculated for different $^{xxx}$Tb-cm09 are reported in Table 4. For all the radioisotopes, the organs receiving the highest absorbed dose are the kidneys, followed by the osteogenic cells for $^{155}$Tb, $^{156m1}$Tb and $^{156m2}$Tb, by the adrenals for $^{154}$Tb and $^{156g}$Tb. $^{156g}$Tb is the radioisotope giving the highest values of absorbed doses, due to its long half-life and the large amount of energy emitted per decay (see Table 1). The absorbed doses due to $^{154}$Tb administration are also higher than those of $^{155}$Tb, because, even if the $^{154}$Tb half-life is much shorter than that of $^{155}$Tb, the energy emitted per decay is about ten times higher. The absorbed doses due to $^{156m1}$Tb and $^{156m2}$Tb are about one order of magnitude lower than those of $^{155}$Tb. The ED after administration of $^{156m1}$Tb-cm09 and $^{156m2}$Tb-cm09 are lower than the one of $^{155}$Tb-cm09, consequently their presence as contaminants in the produced $^{155}$Tb will not increase the dose to the patient. In contrast, when the radiopharmaceutical is labelled with $^{156g}$Tb or $^{154}$Tb, the ED is 5.9 and 2.4 times the ED values of $^{155}$Tb-cm09. Therefore, to guarantee a dose increment per unit of activity administered lower than 10%, the presence of $^{156g}$Tb or $^{154}$Tb as contaminants must be lower than about 2% or 7% respectively.

**Table 3.** Number of nuclear transitions (MBq × h/MBq) in source organs per unit administered activity of $^{xxx}$Tb-cm09 for male ICRP 89 human phantom.

| Organ/Tissue | $^{154g}$Tb-cm09 | $^{155}$Tb-cm09 | $^{156}$Tb-cm09 | $^{156m1}$Tb-cm09 | $^{156m2}$Tb-cm09 |
|---|---|---|---|---|---|
| Heart contents | 0.118 | 0.168 | 0.168 | 0.122 | 0.075 |
| Lung | 0.177 | 0.391 | 0.392 | 0.190 | 0.079 |
| Spleen | 0.014 | 0.048 | 0.048 | 0.016 | 0.005 |
| Kidneys | 0.912 | 2.885 | 2.891 | 1.012 | 0.244 |
| Stomach | 0.020 | 0.043 | 0.044 | 0.022 | 0.008 |
| Left colon | 0.0031 | 0.0072 | 0.0072 | 0.0033 | 0.0014 |
| Small intestine | 0.0146 | 0.0335 | 0.0335 | 0.0156 | 0.0066 |
| Right colon | 0.0062 | 0.0144 | 0.0144 | 0.0067 | 0.0028 |
| Rectum | 0.0031 | 0.0072 | 0.0072 | 0.0033 | 0.0014 |
| Liver | 0.498 | 1.443 | 1.446 | 0.545 | 0.166 |
| Cortical bone | 0.405 | 0.991 | 0.993 | 0.438 | 0.152 |
| Salivary glands | 0.041 | 0.112 | 0.113 | 0.044 | 0.014 |
| Remaining | 3.985 | 8.819 | 8.833 | 4.274 | 1.719 |

**Table 4.** Organ absorbed doses (mGy/MBq) and ED values (mSv/MBq) per unit administered activity calculated for $^{xxx}$Tb-cm09 for male ICRP 89 phantoms.

| Target organ | $^{154g}$Tb-cm09 | $^{155}$Tb-cm09 | $^{156}$Tb-cm09 | $^{156m1}$Tb-cm09 | $^{156m2}$Tb-cm09 |
|---|---|---|---|---|---|
| Adreanals | 1.68E-01 | 6.56E-02 | 4.69E-01 | 6.78E-03 | 1.44E-03 |
| Brain | 2.61E-02 | 9.25E-03 | 5.68E-02 | 1.38E-03 | 1.22E-03 |
| Esophagus | 4.93E-02 | 1.52E-02 | 1.10E-01 | 1.98E-03 | 1.24E-03 |
| Eyes | 2.58E-02 | 9.09E-3 | 5.57E-02 | 1.36E-03 | 1.22E-03 |
| Gallbladder wall | 7.75E-02 | 2.66E-02 | 1.97E-01 | 3.05E-03 | 1.27E-03 |
| Left colon | 5.91E-02 | 2.04E-02 | 1.44E-01 | 2.50E-03 | 1.71E-03 |
| Small intestine | 5.10E-02 | 1.67E-02 | 1.19E-01 | 2.11E-03 | 1.71E-03 |
| Stomach wall | 5.60E-02 | 1.87E-02 | 1.29E-01 | 2.49E-03 | 2.05E-03 |
| Right colon | 5.59E-02 | 1.87E-02 | 1.34E-01 | 2.31E-03 | 1.71E-03 |
| Rectum | 4.13E-02 | 1.33E-02 | 9.11E-02 | 1.80E-03 | 1.70E-03 |
| Heart wall | 6.75E-02 | 2.14E-02 | 1.35E-01 | 3.64E-03 | 4.98E-03 |
| Kidneys | 3.92E-01 | 3.48E-01 | 1.29E00 | 4.27E-02 | 3.99E-02 |
| Liver | 1.01E-01 | 5.06E-02 | 2.77E-01 | 6.10E-03 | 4.74E-03 |
| Lungs | 4.78E-02 | 1.99E-02 | 1.10E-01 | 3.00E-03 | 3.36E-03 |
| Pancreas | 6.14E-02 | 1.97E-02 | 1.46E-01 | 2.37E-03 | 1.22E-03 |
| Prostate | 4.15E-02 | 1.23E-02 | 9.27E-02 | 1.55E-03 | 1.22E-03 |
| Salivary Glands | 6.42E-02 | 4.67E-02 | 1.74E-01 | 6.57E-03 | 8.32E-03 |
| Red Marrow | 4.11E-02 | 1.15E-02 | 9.39E-02 | 1.41E-03 | 1.03E-03 |
| Osteogenic cells | 6.00E-02 | 8.61E-02 | 1.75E-01 | 1.78E-02 | 1.04E-02 |
| Spleen | 7.72E-02 | 3.27E-02 | 2.12E-01 | 3.53E-03 | 1.75E-03 |
| Testes | 2.92E-02 | 8.58E-03 | 6.09E-02 | 1.20E-03 | 1.21E-03 |
| Thymus | 4.31E-02 | 1.22E-02 | 8.86E-02 | 1.75E-03 | 1.24E-03 |
| Thyroid | 3.55E-02 | 1.10E-02 | 7.60E-02 | 1.54E-03 | 1.22E-03 |
| Urinary bladder wall | 3.82E-02 | 1.12E-02 | 8.24E-02 | 1.48E-03 | 1.22E-03 |
| Total body | 3.57E-02 | 1.36E-02 | 8.13E-02 | 1.91E-03 | 1.74E-03 |
| Effective dose | 4.44E-02 | 1.86E-02 | 1.09E-01 | 2.47E-03 | 2.06E-03 |

**The dose increase due to $^{155}$Tb contaminants obtained with different levels of target enrichment**

The DI due to $^{155}$Tb-cm09 obtained with different levels of target enrichment is plotted in Fig. 8 vs. time *t* post-production. The DI is well above 25% for the 91.9% target enrichment and this confirms that the isotopic contamination of that target is inadequate for medical purposes. The ideal case of a $^{155}$Gd target with 100% enrichment is described by the solid red line. Here, the RNP is very close to 100% and the DI is negligible. Quite interesting are the dashed and dot-dashed curves obtained with 99% and 98% enrichment. The former reaches a 5% DI, while the latter remains

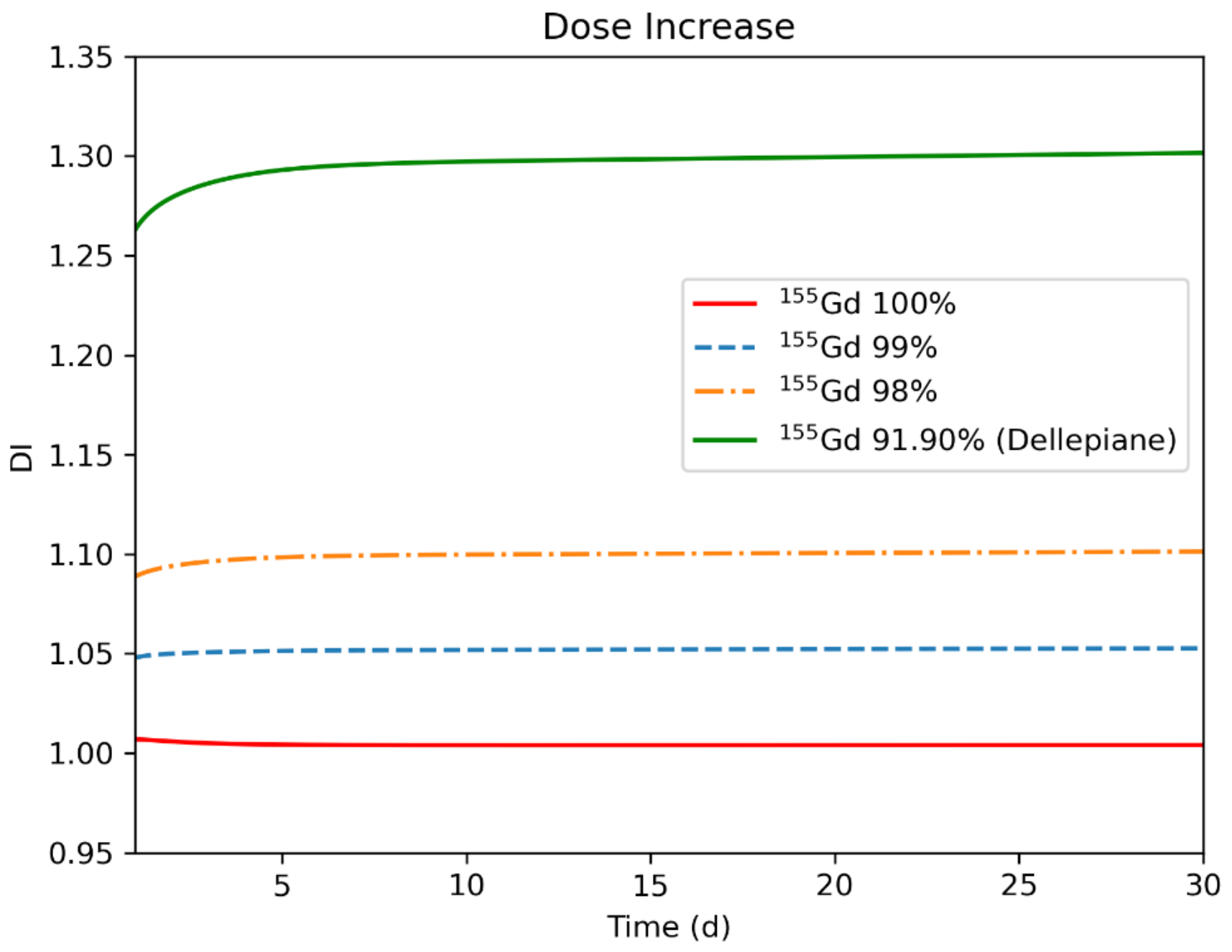

within the 10% limit.

**Fig. 8** Dose increase for $^{155}$Tb-cm09 radiopharmaceutical labeled at different times after the end of irradiation of different $^{155}$Gd-enriched targets.

**Impact on the image quality**

The imaging quality of $^{155}$Tb produced with the different enriched $^{155}$Gd targets was assessed at the EoB and 96 h later by considering the $^{156g}$Tb and $^{154g}$Tb yields reported in Table 2. The metastable states $^{156m1}$Tb, $^{156m2}$Tb and $^{154m1}$Tb were disregarded in the simulations because of their lower energy gamma ray emission.

A typical gamma-ray spectrum is shown in Fig. 9. It has been simulated for $^{155}$Tb produced at the EoB from the irradiation of 100% enriched $^{155}$Gd target. The gamma rays with the energies of 86.55 keV and 105.318 keV emitted from $^{155}$Tb and the low intensity 88.97 keV gamma ray emitted by $^{156g}$Tb are situated close with respect to the energy resolution of the imaging system and form a compound peak with a barycenter of 88.5 keV. Other four peaks from $^{155}$Tb with energies of 148.64 keV, 161.29 keV, 163.28 keV and 180.08 keV form a less intense compound peak with a barycenter of 167 keV. A smaller $^{155}$Tb peak at 262 keV (5.3%) is exposed to the higher energy gamma rays from the isotopes of $^{154g}$Tb and $^{156g}$Tb only. For imaging purposes, the compound peak with the energy of 88.5 keV is most convenient, for its higher intensity, though the other two peaks at 167 keV and 262 keV can also be used for imaging.

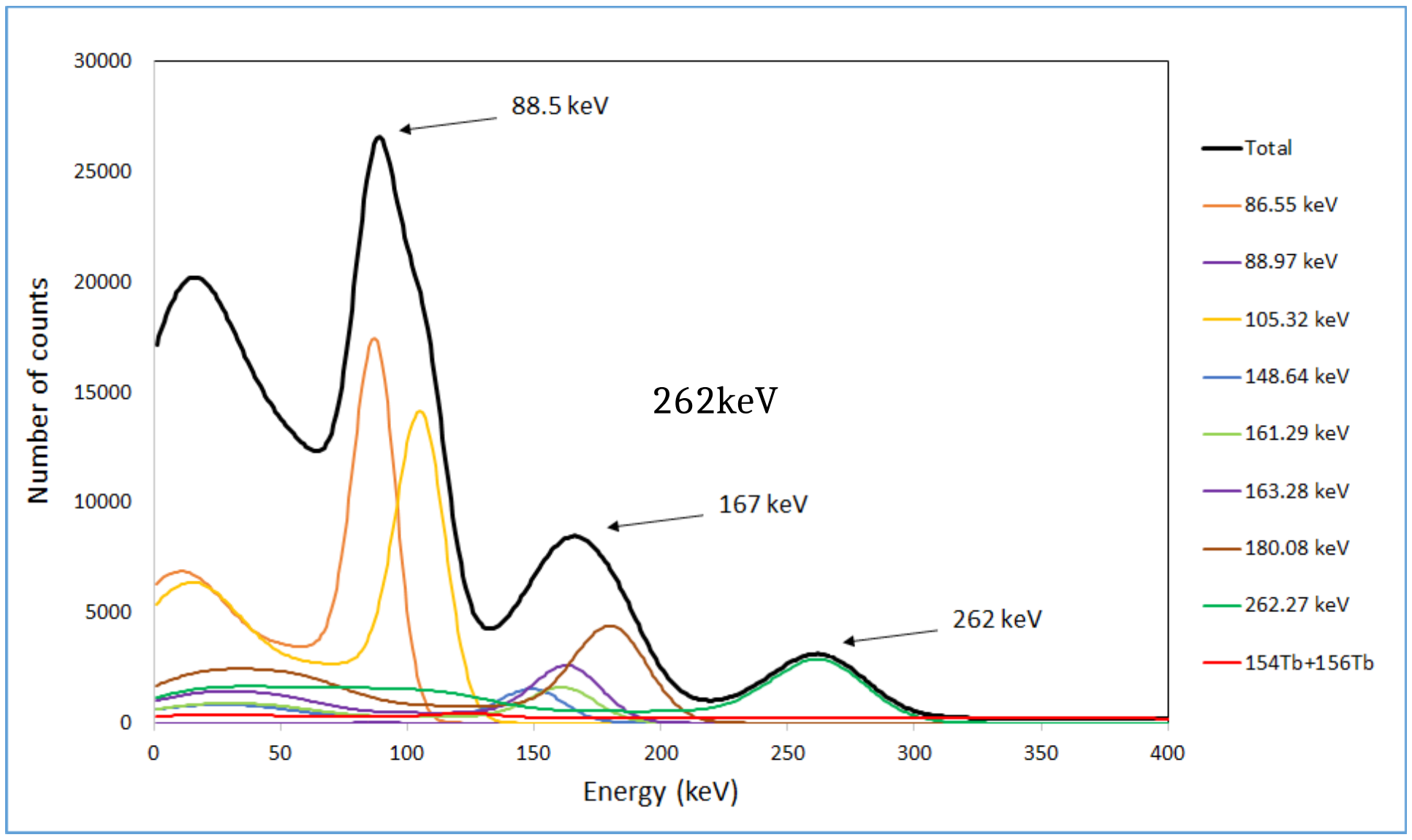

**Fig 9.** Simulated spectrum for $^{155}$Tb, obtained immediately after the proton bombardment of 100% enriched $^{155}$Gd target. The predicted observable spectrum is presented with a thicker black line.

Table 5 shows the calculated Compton-to-peak ratio for the three principal peaks of $^{155}$Tb with respect to the gamma-ray background, generated by the contaminant isotopes $^{156g}$Tb and $^{154g}$Tb. The analysis of the data reveals that in all cases the level of noises in the gamma ray images reconstructed using a 10% energy window around the barycenter of the peaks does not exceed 27%. For the intense peak at 88.5 keV the Compton-to-peak ratio remains within the interval of 19%– 23% for all kind of targets immediately after the EoB and no significant improvement could be achieved 96 h later. Somewhat lower Compton-to-peak values below 20% are obtained for the compound peak at 167 keV. This suggests its use as a single peak for image reconstruction in some particular cases. The lower intensity of the peak at 262 keV makes the ratio more affected by the presence of the contaminant nuclides $^{156g}$Tb and $^{154g}$Tb though its maximum value still remains below 27%. However, it is worth noting that the present image quality estimation is made for point-like gamma-ray sources in air, hence the values quoted refer to the imaging system only and do not take into account Compton scattering inside the tissues. To estimate this contribution for the case of small animals imaging, a water phantom was used in our laboratory, obtaining a noise increase level smaller than 5%.

**Table 5.** Compton-to-peak ratio calculated from the simulation of $^{155}$Tb gamma-ray spectra, obtained for 100%, 99% and 98% enrichment of $^{155}$Gd targets, at the EoB and 96 h later.

| | Compton-to-peak ratio EoB | | | Compton-to-peak ratio 96 h | | |
|---|---|---|---|---|---|---|
| Peak energy (keV) | 100% $^{155}$Gd | 99% $^{155}$Gd | 98% $^{155}$Gd | 100% $^{155}$Gd | 99% $^{155}$Gd | 98% $^{155}$Gd |
| 88.5 | 19.44 % | 20.99 % | 22.59 % | 18.89 % | 21.52 % | 22.61 % |
| 167 | 10.24 % | 14.54 % | 19.06 % | 8.91 % | 15.95 % | 19.32 % |
| 262 | 5.84 % | 16.12 % | 26.9 % | 1.71 % | 19.8 % | 26.55 % |

## Discussion

The use of low-energy proton beams on $^{155}$Gd-enriched target represents a promising possibility for the production of $^{155}$Tb, however the isotopic purity of the target strongly influences the amount of co-produced radioisotopes. The most problematic contaminant is $^{156g}$Tb, because its half-life is comparable to that of $^{155}$Tb, and its $\gamma$ emissions have severe impact on the dose released and on the quality of the SPECT images. For these reasons, its production should be kept as low as possible. The co-production of $^{156g}$Tb can be minimized by limiting the $^{156}$Gd content in the target. In particular, a RNP higher than 98.5% can be obtained starting from about 30 h after the EoB with a $^{156}$Gd content lower than 1%, slowly approaching a maximum value of 99%. However, the 99% value is not an established limit and a smaller one can be tolerated if the DI due to the presence of contaminant radioisotopes is low. By assuming a 10% limit in DI as an acceptable condition for the contamination of a production route (31), the maximum content of $^{156}$Gd in the target could be increased to 2%. It is worth to note that in this case the DI does not exceed the 10% value in the entire time range shown in Fig. 8, namely from 0 to 30 days. Therefore the available activity can be utilized much earlier with respect to the timing when the RNP is close to 98%, and specifically, the product can be available right after the time needed to perform a radiochemical purification (9).

Enriched $^{155}$Gd is currently commercially available from various companies (ISOFLEX, CIL, AMT, etc.) with $^{155}$Gd purity greater than 90%, but with a $^{156}$Gd component larger than 2%, not yet sufficient for the production of $^{155}$Tb with the necessary purity for medical application. Gd-isotopes separation is performed commercially using the "Calutron method", namely using a very large mass spectrometer for electro-magnetic separation. Since this technique is very energy inefficient, the prices for the enriched Gd material are high. However, it is possible to tune the Calutron to produce highly enriched $^{155}$Gd with isotopic enrichment >98.0%, or with $^{156}$Gd content <2%. To make a producer interested in changing the operational plan implies an economical return and profitability which presently is not yet expected by the isotope-producer companies (Allan Pashkovski, Managing Director, Isoflex, Private communications, 6th July, 2023). A scenario of a diffuse worldwide production of $^{155}$Tb by hospital cyclotrons may be much more favorable for the expectations of economical return by the producer companies.

The targeted theranostic agents currently used in nuclear medicine are based on biological structures such as peptides, monoclonal antibodies and their fragments, coupled to α and β emitters radionuclides. However, due to their nature, these agents exhibit maximum absorption time range from several hours to a few days after injection. $^{111}$In is the only radionuclide currently used in the clinic to perform SPECT imaging of these radiocomplexes since its half-life (2.8 d) is long enough for image acquisition even several days after the radiocomplex administration. Therefore, to evaluate the $^{155}$Tb potential as matched pair for targeted theranostic agents labelled with

$^{149}$Tb and $^{161}$Tb, it is interesting to compare its imaging and dosimetric properties with those of $^{111}$In.

Imaging simulations were performed for pure $^{111}$In using the main photon emissions at energies of 171 keV (91%) and 245 KeV (94%) to compare its Compton-to-peak ratio with those of $^{155}$Tb. The Compton-to-peak ratio of the peak at 171 keV, the closest in energy to the 88.5 keV and 167 keV peaks of $^{155}$Tb, turned out to be 22.6%, which is slightly larger than the values obtained for $^{155}$Tb, although they are comparable. These results are supported by the excellent imaging properties of $^{155}$Tb reported by Favaretto (9), where a spatial resolution up to 1 mm was obtained from a Derenzo phantom. Moreover, Muller *et al.* (5) published a comparison between SPECT images of Derenzo phantoms filled with solutions of pure $^{155}$Tb (2.6 MBq) and $^{111}$In (4 MBq), demonstrating comparable spatial resolution and the ability to produce equal images with lower activity of $^{155}$Tb.

Dosimetric properties of $^{111}$In-labelled cm09 were estimated with OLINDA by assuming the same biodistribution of Tb-cm09 compound. The calculated ED of $^{111}$In-cm09 was 0.0217 mSv/MBq, very similar to that of $^{155}$Tb-cm09 (0.0186 mSv/MBq). These results are a consequence of the fact that, despite the half-life of $^{155}$Tb is almost 2 times longer than that of $^{111}$In, its total energy emission per decay is about 2 times lower compared to $^{111}$In (0.4409 MeV/nt). It should therefore be expected that the imaging and dosimetric properties of

$^{155}$Tb-radiocomplexes be comparable and even better of those of $^{111}$In ones.

**Conclusions**

$^{155}$Tb can be produced with a quality suitable for medical applications using low-energy proton beams and $^{155}$Gd-enriched targets, if the $^{156}$Gd impurity content does not exceed 2%. Under these conditions, the dose increase due to the presence of contaminant radioisotopes remains below the 10% limit and good quality images, comparable to those of $^{111}$In, are guaranteed.

## List of abbreviations

DI Dose increase

ED Effective dose

EoB End of bombardment

GDH Geometry dependent hybrid

LD Level-density

OLINDA Organ Level Internal Dose Assessment

PE Pre-equilibrium

PET Positron emission tomography

RNP Radionucludic purity

SPECT Single photon emission computed tomography

## Declarations

### Ethics approval and consent to participate

Not applicable

### Consent for publication

Not applicable

### Availability of data and material

The datasets used and/or analysed during the current study are available from the corresponding author on reasonable request.

### Competing interests

The authors declare that they have no competing interests.

### Funding

This work was supported by the REMIX-CSN5 research program (2021-2023), funded by Italian National Institute of Nuclear Physics (INFN) as part of the activities of the LARAMED project of the INFN-Legnaro National Laboratories. The work of LC and FB was also funded by NUCSYS-CSN4 INFN research program.

## Authors' contributions

LC, LDN, and LMA contributed to the study conception and design. FB and LC performed cross section calculation and assessed yields, isotopic and radionuclidic purity for targets with different compositions. LDN, LMA, and FB obtained biokinetic curves and performed dosimetric studies. NU evaluated the impact of impurities on image quality. The manuscript was written, read and approved by all authors.

## Acknowledgements

The authors thank Allan Pashkovski for private communications.

One of the authors (N. Uzunov) would like to acknowledge the support received from the Abdus Salam International Center for Theoretical Physics Trieste, Italy, and in particular the Program for Training and Research in Italian Laboratories (TRIL).